\documentclass[preprint,aps,showpacs]{revtex4}

\usepackage{graphics}

\begin{document}

\title{A Novel Quasi-Exactly Solvable Model with Total Transmission Modes}

\author{Hing-Tong Cho}
  \email{htcho@mail.tku.edu.tw}
\author{Choon-Lin Ho}
  \email{hcl@mail.tku.edu.tw}
\affiliation{Department of Physics, Tamkang University, Tamsui,
Taipei, Taiwan, Republic of China}

\date{\today}

\begin{abstract}

In this paper we present a novel quasi-exactly solvable model with
symmetric inverted potentials which are unbounded from below. The
quasi-exactly solvable states are shown to be total transmission
(or reflectionless) modes. From these modes even and odd
wavefunctions can be constructed which are normalizable and
flux-zero. Under the procedure of self-adjoint extension, a
discrete spectrum of bound states can be obtained for these
inverted potentials and the solvable part of the spectrum is the
quasi-exactly solvable states we have discovered.

\end{abstract}


\maketitle

We present here a novel quasi-exactly solvable (QES) model with an
inverted potential which is unbounded from below. Quasi-exact
solvability means that part of the energy eigenvalues and the
corresponding eigenfunctions can be solved algebraically
\cite{turbiner, shifman, ushveridze,gonzalez,ulyanov}. This can be
considered as an extension to the exactly solvable quantum
mechanical systems.  There is in fact a Lie-algebraic structure
behind most QES systems. Particularly, all one-dimensional QES
systems have been classified based on the $sl(2)$ algebra
\cite{turbiner,gonzalez}. These $sl(2)$-based QES systems can also
be interpreted as appropriate spin systems \cite{ulyanov}. As
exactly solvable systems are rather rare, the discovery of QES
models in the late 80's has thus greatly enlarged the scope of
spectral problems in quantum physics.  Moreover, they have been
found to have applications in quantum field theory (conformal
theory \cite{morozov} in particular) and condensed matter physics
\cite{wiegmann,hatsugai} in recent years.

In the course of looking for QES spectra involving inverted
potentials  \cite{cho}, we found a new QES potential,
\begin{equation}
V(x)=-\frac{b^{2}}{4}{\rm
sinh}^{2}x-\left(n^{2}-\frac{1}{4}\right){\rm
sech}^{2}x,\label{potential}
\end{equation}
where $b>0$ is a real parameter and $n=1,2,3,\dots$. To see how we
obtain this QES potential, we start with a $sl(2)$-based QES
Hamiltonian $H=-d^2/dx^2 +V(x)$ ($-\infty <x<\infty$) with this
property.  It belongs to the ``case 1" in \cite{gonzalez}
involving hyperbolic functions. The general form of $V(x)$ in case
1 is
\begin{eqnarray}
V(x)&=& A {\rm sinh}^2\sqrt{\nu} x+ B{\rm
sinh}\sqrt{\nu} x\nonumber\\
&+& C {\rm tanh}\sqrt{\nu} x{\rm sech}\sqrt{\nu} x + D {\rm
sech}^2\sqrt{\nu} x,
\end{eqnarray}
where $\nu$ is a real scale factor, and $A,~B,~C$ and $D$ are real
constants which are algebraically constrained. In \cite{gonzalez}
it has been concluded that if one requires the eigenfunctions to
be normalizable, only the exactly solvable Scarf II potentials,
with $A=B=0$, are left in this case (see also, for example,
Table~4.1 in \cite{cooper}).  The system defined by
Eq.~(\ref{potential}) was not considered in \cite{gonzalez}
because $V(x)$ above is bottomless and is thus singular at plus
and minus infinity.

However, as we discuss in details below, normalizable QES bound
states are indeed possible in this system. This corresponds to the
case with $B=C=0$. More precisely, the potential takes the form
(we rescale $\nu$ to unity without loss of generality) of
Eq.~(\ref{potential}). It is a two-parameter symmetric potential.
As $x\rightarrow\pm\infty$, $V(x)\rightarrow -\infty$, so it is a
bottomless potential. For $b<2\sqrt{(n^{2}-1/4)}$, the potential
is like an inverted double well, while for $b\geq
2\sqrt{(n^{2}-1/4)}$, it is like an inverted oscillator with one
maximum. Potentials like these are mostly thought not to admit
normalizable states. However, we have found that $V(x)$ not only
supports normalizable QES states but it also exhibits very
peculiar features.

The wavefunction of the system defined by the potential in
Eq.~(\ref{potential}) takes the general form
$\psi=\exp(-g(x))\phi(x)$, where $g(x)=ib~{\sinh} x/2 +(n-1/2)\ln
{\rm cosh}x$.  The function $\phi(x)$ satisfies a Schr\"odinger
equation with a transformed Hamiltonian $H_g=e^g H e^{-g}$ given
by
\begin{eqnarray}
H_g=-\left(z^2+1\right)\frac{d^2}{dz^2}
+\left(ibz^2+2\left(n-1\right)z+ib\right)\frac{d}{dz}
-ib\left(n-1\right)z+\frac{b^2}{4}-\left(n-\frac{1}{2}\right)^2,
\label{Hg}
\end{eqnarray}
where $z(x)\equiv \sinh x$.  It is easily shown that
Eq.~(\ref{Hg}) can be expressed as a quadratic combination of the
generators $J^a$ of the $sl(2)$ Lie algebra with an
$n$-dimensional representation:
\begin{eqnarray}
H_g=-J^+J^-  -J^-J^- + ib J^+ +ib J^- + \left(n-1\right) J^0
+\frac{b^2}{4} - \frac{1}{2}\left(n^2-\frac{1}{2}\right).
\label{Hg2}
\end{eqnarray}
Here the generators $J^a$ of the $sl(2)$ Lie algebra  take the
differential forms: $ J^+ = z^2 d/dz - (n-1)z~,~ J^0=z
d/dz-(n-1)/2~,~J^-=d/dz$ ($n=1,2,\ldots$). Within this finite
dimensional Hilbert space the Hamiltonian $H_g$ can be
diagonalized, and therefore a finite number (which is just $n$) of
eigenstates are solvable. The system described by $H$ is therefore
QES.  Eq.~(\ref{Hg2}) can thus be considered as the spin
Hamiltonian \cite{ulyanov} corresponding to Eq.~(\ref{potential})
which reveals the underlying $sl(2)$ structure of the system.

\begin{figure}[!]
\includegraphics{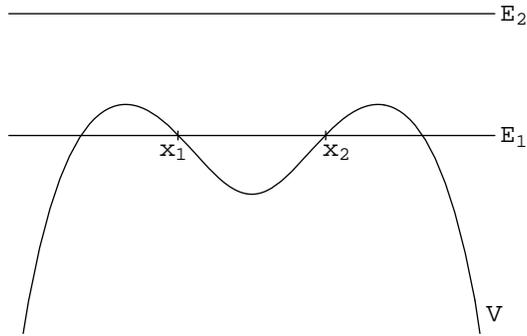}
\caption{\label{fig1}QES energy levels of the potential $V(x)$
with $b=1$ and $n=2$.}
\end{figure}

To make the following discussion more concrete, we first take
$b=1$. In this case, we have an inverted double well. Although the
general features of the problem will not depend on this choice, we
shall indicate how the details vary with the value of $b$ later.
It is a straightforward procedure
\cite{turbiner,shifman,ushveridze,gonzalez} to obtain the QES
spectra of $V(x)$ for various values of $n$. The number of
distinct values of the QES energies is just $n$.  For $n=1$, the
QES states have energy $E^{(n=1)}_{1}=0$, and the corresponding
wavefunctions are
\begin{equation}
\psi^{(n=1)}_{1(R,L)}=({\rm cosh}x)^{-1/2}e^{\pm\frac{i}{2}{\rm
sinh}x}.\label{n1TT}
\end{equation}
Here $R(L)$ in the subscript represents right-(left-)moving mode.
From their probability fluxes,
\begin{eqnarray}
&&j^{(n=1)}_{1(R,L)}\nonumber\\
&\equiv&
i\left[\left(\frac{d\psi^{(n=1)*}_{1(R,L)}}{dx}\right)
\psi^{(n=1)}_{1(R,L)}
-\psi^{(n=1)*}_{1(R,L)}\left(\frac{d\psi^{(n=1)}_{1(R,L)}}{dx}\right)
\right]\nonumber\\ &=&\pm 1,
\end{eqnarray}
we see that these are scattering states with constant (no
reflection) unit fluxes. Hence our QES states represent right- and
left-moving total transmission (TT) modes.  This is also true for
other values of $n$. The QES spectra for $n$ up to 5 are tabulated
in Table I. For $n=2$, the corresponding TT mode wavefunctions are
\begin{eqnarray}
\psi^{(n=2)}_{1(R,L)}&=&\frac{e^{\pm\frac{i}{2}{\rm
sinh}x}}{(1-\sqrt{2})({\rm cosh}x)^{3/2}}\left[1\pm
i(1-\sqrt{2}){\rm sinh}x\right],\\
\psi^{(n=2)}_{2(R,L)}&=&\frac{e^{\pm\frac{i}{2}{\rm
sinh}x}}{(1+\sqrt{2})({\rm cosh}x)^{3/2}}\left[1\pm
i(1+\sqrt{2}){\rm sinh}x\right].
\end{eqnarray}
All of them have unit fluxes. Here 1 and 2 in the subscript
represent the first and the second QES modes with $E_{1}<E_{2}$.
It is interesting to see that the state $\psi^{(n=2)}_{1}$ has
energy $E_{1}$ which is under the peaks of the potential, that is,
in the valley as shown in Fig.~\ref{fig1}. There is no reflection
even though the scattering state tunnels through two barriers.
With other values of $n$, one again finds total transmission
tunneling states. Their energies are given in parentheses in
Table~\ref{table1}.

\begin{table}[!]
\caption{\label{table1}QES energy spectra for the potential in
Eq.~(\ref{potential}) with $b=1$ and $n=1$ to 5. The energy levels
in the valley of the potential are given in parentheses.}
\begin{ruledtabular}
\begin{tabular}{cccccc}
n & $E_{1}$ & $E_{2}$ & $E_{3}$ & $E_{4}$ & $E_{5}$ \\ \hline 1 &
0 & & & &
\\ 2 & $(-2.4)$ & 0.4 & & &
\\ 3 & $(-6.340)$ & $-2.622$ & 0.962 & &
\\ 4 & $(-12.301)$ & $(-6.523)$ & $-2.760$ & 1.585 &
\\ 5 & $(-20.286)$ & $(-12.405)$ & $(-6.756)$ & $-2.806$ & 2.253 \\
\end{tabular}
\end{ruledtabular}
\end{table}

For the inverted potential that we are dealing with, it is obvious
that the spectrum for the TT modes is discrete and bounded from
below. To understand this spectrum in a more quantitatively way,
and to see how the QES spectrum fits into it, we use the WKB
approximation to estimate the energies of the TT modes. For
energies under the bottom of the valley, that is, when there are
only two turning points, no TT modes are possible. This is
consistent with the usual understanding that after tunneling the
transmitted flux is much small than the incident flux. The
energies are thus bounded from below by the value at the bottom of
the valley. For $n=1$, it is $\sim-3.75$. When the energy is in
the valley of the potential, there are four turning points.
Following the usual WKB procedure \cite{bender} of matching the
wavefunctions from the incident side to the transmitted side
through all the turning points, one arrives at the quantization
rule, in the lowest order,
\begin{equation}
\int_{x_{1}}^{x_{2}}dx\sqrt{E-V(x)}=\frac{\pi}{2}\ ,\
\frac{3\pi}{2}\ ,\ \frac{5\pi}{2}\ , \dots\label{rule}
\end{equation}
where $x_{1}$ and $x_{2}$ are the turning points in the valley as
shown in Fig.~\ref{fig1}. The energies which satisfy this
quantization rule for various values of $n$ are listed in
Table~\ref{table2}. We have also shown the percentage errors of
these values in comparison with the exact values in
Table~\ref{table1}. The errors are smaller when the levels are
nearer to the bottom of the valley. The matching of these WKB
levels with the QES states indicates that the QES states indeed
constitute the lowest energy states in the TT spectra. Above the
peaks of the potentials the WKB states are all TT modes in the
lowest order. To estimate the energies in these cases, one has to
go to higher WKB orders which we shall not pursue here.

\begin{table}[!]
\caption{\label{table2}Energies calculated using the WKB
approximation for the TT modes in the valley of the potentials.
The percentage errors from comparing with the exact values in
Table~I are given in parentheses.}
\begin{ruledtabular}
\begin{tabular}{cccc}
n & $E_{1}$ & $E_{2}$ & $E_{3}$ \\ \hline  2 & $-2.173$ (10\%) & &
\\ 3 & $-6.099$ (4\%) & &
\\ 4 & $-12.070$ (2\%) & $-6.299$ (3\%) &
\\ 5 & $-20.055$ (1\%) & $-12.208$ (1.6\%) & $-6.527$ (3.4\%) \\
\end{tabular}
\end{ruledtabular}
\end{table}

Next, we would like to discuss how the considerations above will
be modified when $b$ is varied. We have mentioned that at
$b=2\sqrt{(n^{2}-1/4)}$ the potential changes from an inverted
double well to an inverted oscillator potential when $b$ is
increased. Regarding the QES spectra there are other interesting
values of $b$. When $b$ is very small but non-zero, all the QES
states lie in the valley. Actually the maximum number $n_{max}$ of
TT states in the valley can be estimated using the quantization
rule in Eq.~(\ref{rule}) by taking $x_{1}$ and $x_{2}$ to be the
locations of the peaks, $\pm x_{peak}$ and with $b\rightarrow 0$.
As $b\rightarrow 0$, $V(x)\sim-(n^{2}-1/4){\rm sech}^{2}x$,
$x_{peak}\rightarrow\infty$, and $V(x_{peak})\rightarrow 0$. The
integral in the quantization rule becomes
\begin{eqnarray}
\int_{-\infty}^{\infty}dx\ \sqrt{\left(n^{2}-\frac{1}{4}\right)}\
{\rm sech}x&=&\sqrt{\left(n^{2}-\frac{1}{4}\right)}\
\pi\nonumber\\ &\geq& \left(n_{max}-\frac{1}{2}\right)\pi .
\end{eqnarray}
Hence, $n_{max}=n$, that is, when $b\rightarrow 0$ all the TT
states in the valley are QES states. As $b$ is increased, the
height of the peaks decreases while the energies of the QES states
increase. At one point the highest QES level starts to leave the
valley. For example, for $n=1$, this happens when
$b=1/2\sqrt{3}\sim0.289$. When $b$ is increased further, the peaks
merge into one at $b=\sqrt{3}\sim1.732$. For $n=2$, the highest
level $E_{2}$ leaves the valley when
$b=(5\sqrt{15}-2\sqrt{69})/22\sim0.125$. When $b$ is increased
further, the lower level $E_{1}$ will leave the valley at
$b=(5\sqrt{15}+2\sqrt{69})/22\sim1.635$. The peaks merge when
$b=\sqrt{15}\sim3.873$. Similar situation happens for other values
of $n$.

Up to now we have indicated that the QES states are TT scattering
states. However, one can in fact construct bound states from these
scattering states. This peculiar situation that scattering states
and bound states share the same energies is very much related to
the asymptotic behavior of the potential. In particular, as long
as the potential goes to $-\infty$ faster than $-|x|^{s}$ with
$s>2$ as $x$ goes to $\pm\infty$, normalizable states can be
constructed out of the scattering states. This can be understood
in the following intuitive argument. Classically we would have
runaway solutions with the speed of the particle growing as $|x|$
increases for this kind of potentials. However, the quantum
mechanical probability density can be viewed roughly as inversely
proportional to the speed of the particle. If the speed grows fast
enough as stated above, the probability density will be suppressed
to such an extend that the wavefunction becomes normalizable.

In our model we can construct these normalizable states as
follows. For $n=1$, from the TT wavefunctions in Eq.~(\ref{n1TT}),
we obtain the even ($+$) and odd ($-$) wavefunctions
\begin{eqnarray}
\psi^{(n=1)}_{1\pm}&\sim&\psi^{(n=1)}_{1R}\pm\psi^{(n=1)}_{1L}\nonumber\\
&\sim&\left\{
\begin{array}{l}
({\rm cosh}x)^{-1/2}{\rm cos}(\frac{1}{2}{\rm sinh}x),\\ ({\rm
cosh}x)^{-1/2}{\rm sin}(\frac{1}{2}{\rm sinh}x).
\end{array}
\right.\label{n1bound}
\end{eqnarray}
$\psi^{(n=1)}_{1\pm}$ have zero fluxes by construction.
Furthermore, they are normalizable since
\begin{eqnarray}
\int^{\infty}_{-\infty}dx|\psi^{(n=1)}_{1\pm}|^{2}&\sim&
\int^{\infty}_{-\infty}dx({\rm cosh}x)^{-1}{\rm
cos}^{2}(\frac{1}{2}{\rm sinh}x)\nonumber\\
&\sim&\frac{\pi}{2}(1+e^{-1}),
\end{eqnarray}
which is finite. These are normalizable states with zero fluxes.
Similar normalizable states can also be constructed for other
values of $n$. For example, for $n=2$, they are
\begin{eqnarray}
\psi_{(1,2)+}^{(n=2)}&\sim&({\rm cosh}x)^{-3/2}\Big[{\rm
cos}\left(\frac{1}{2}{\rm sinh}x\right)\nonumber\\ &&\ \ \ \ \
-(1\mp\sqrt{2}){\rm sinh}x\ {\rm sin}\left(\frac{1}{2}{\rm
sinh}x\right)\Big],\\ \psi_{(1,2)-}^{(n=2)}&\sim&({\rm
cosh}x)^{-3/2}\Big[{\rm sin}\left(\frac{1}{2}{\rm
sinh}x\right)\nonumber\\ &&\ \ \ \ \ +(1\mp\sqrt{2}){\rm sinh}x\
{\rm cos}\left(\frac{1}{2}{\rm sinh}x\right)\Big].
\end{eqnarray}
In fact, normalizable zero-flux states can be found for all the
QES energies.

Do these normalizable zero-flux states constitute a bound state
energy spectrum for the hamiltonian operator? We consider this
question by examining the Wronskians between various normalizable
zero-flux states that we have constructed. For example, for $n=2$,
we have
\begin{eqnarray}
\left.W[\psi_{1+}^{(n=2)},\psi_{1-}^{(n=2)}]\right|_{x\rightarrow\pm\infty}
&=&\left[\left(\frac{d\psi_{1+}^{(n=2)}}{dx}\right)\psi_{1-}^{(n=2)}
-\psi_{1+}^{(n=2)}\left(\frac{d\psi_{1-}^{(n=2)}}{dx}\right)
\right]_{x\rightarrow\pm\infty}\nonumber\\
&=&\sqrt{2}-\frac{3}{2},\nonumber\\
\left.W[\psi_{2+}^{(n=2)},\psi_{2-}^{(n=2)}]\right|_{x\rightarrow\pm\infty}
&=&-\sqrt{2}-\frac{3}{2},\nonumber\\
\left.W[\psi_{1+}^{(n=2)},\psi_{2-}^{(n=2)}]\right|_{x\rightarrow\pm\infty}
&=&\frac{1}{2},\nonumber\\
\left.W[\psi_{1-}^{(n=2)},\psi_{2+}^{(n=2)}]\right|_{x\rightarrow\pm\infty}
&=&-\frac{1}{2},\nonumber\\
\left.W[\psi_{1+}^{(n=2)},\psi_{2+}^{(n=2)}]\right|_{x\rightarrow\pm\infty}
&=&\left.W[\psi_{1-}^{(n=2)},\psi_{2-}^{(n=2)}]\right|_{x\rightarrow\pm\infty}
=0.\label{n2wron}
\end{eqnarray}
We see that the Wronskians between these normalizable flux-zero
states do not vanish asymptotically as usually happen for bound
states. The reason for this is that, as we have discussed above,
classically the speed of the particle grows as
$x\rightarrow\pm\infty$, the time it needs to get to infinity is
actually finite. Quantum mechanically a wave packet will then
disappear in finite time, and the probability is lost. This is
manifested in the non-vanishing asymptotic values for the
Wronskian, and the hamiltonian operator is thus not necessarily
hermitian. To render the hamiltonian hermitian, one must restrict
the domain on which the hamiltonian acts in the Hilbert space.
Mathematically, this procedure is called a self-adjoint extension
\cite{titchmarsh,reed,carreau,feinberg}.

In \cite{cho2}, we have discussed the problem of self-adjoint
extensions for a general symmetric inverted potential. Here we
just adopt the scheme with the requirement that the Wronskians
between any states in the spectrum approach the same limit as
$x\rightarrow\pm\infty$. All the Wronskians in Eq.~(\ref{n2wron})
satisfy this requirement, and all of them have
\begin{equation}
\left.W[\psi_{i},\psi_{j}]\right|_{-\infty}^{\infty}=0.
\end{equation}
Hence, the boundary terms cancel and the hamiltonian can be proved
to be hermitian with respect to the normalizable flux-zero states
constructed from the QES total transmission modes. In fact, under
this self-adjoint extension scheme \cite{cho2}, one obtains a
discrete spectrum of bound states. The even and odd wavefunctions
constructed from the TT modes, including the quasi-exact solvable
ones, constitute part of this spectrum.

One more curious aspect is that the bound states we have found
above are doubly degenerate. This seems to contradict the usual
belief that in one dimension bound states are non-degenerate. One
can indeed prove that for states with the same energy their
Wronskian is a constant. For the usual bound states this constant
is zero by looking at the asymptotic behavior of the bound state
wavefunctions. However, the bound states we have obtained have
very peculiar asymptotic behavior and this constant is not zero,
as shown in Eq~(\ref{n2wron}), even though the wavefunctions
themselves are normalizable. Similar situation happens with the
potential,
\begin{equation}
V(x)=-A_{1}{\rm
cosh}^{2\nu}x-\frac{\nu}{2}\left(\frac{\nu}{2}+1\right){\rm
sech}^{2}x,\label{kkpotential}
\end{equation}
proposed by Koley and Kar in a recent paper \cite{koley} in
relation with fermion localization on branes \cite{koley2}. They
have found a pair of exact energy states,
\begin{eqnarray}
\psi_{1}(x)&\sim&\frac{1}{({\rm
cosh}x)^{\nu/2}}\cos\left[\sqrt{A_{1}}\int({\rm
cosh}x)^{\nu}dx\right],\nonumber\\ \psi_{2}(x)&\sim&\frac{1}{({\rm
cosh}x)^{\nu/2}}\sin\left[\sqrt{A_{1}}\int({\rm
cosh}x)^{\nu}dx\right].\label{kkstate}
\end{eqnarray}
For $\nu=1$, the potentials in Eq.~(\ref{kkpotential}) and in
Eq.~(\ref{potential}) coincide. The pair of states in
Eq.~(\ref{kkstate}) is the same as $\psi_{1\pm}^{(n=1)}$.
Actually, for all positive values of $\nu$, this pair of states
can be shown to be the even and odd wavefunctions constructed from
a total transmission mode. Since the asymptotic value of these two
states,
\begin{equation}
\left.W[\psi_{1},\psi_{2}]\right|_{x\rightarrow\pm\infty}
=-\sqrt{A_{1}},
\end{equation}
which does not vanish, the two states are therefore degenerate.

In summary, we have discovered a novel QES model involving a
symmetric inverted potential which is unbounded from below. Under
the self-adjoint extension procedure \cite{cho2}, a discrete bound
state spectrum is obtained for this potential. The QES states
constitute part of the spectrum which is exactly solvable. This
result should be relevant to different physical phenomena
involving inverted potentials. The problem of fermion localization
on branes as we have discussed briefly above is one example. Other
examples can be found in inflation \cite{guth} and quintessence
\cite{caldwell} models in which one considers the dynamics of
quantum fields rolling down inverted potentials. With the QES
model, one has a set of exact wavefunctions which should be useful
in detail understanding of a lot of the physics behind these
phenomena.

This work was supported in part by the National Science Council of
the Republic of China under the Grants NSC 96-2112-M-032-006-MY3
(H.T.C.) and NSC 96-2112-M-032-007-MY3 (C.L.H.). The authors would
also like to thank the National Center for Theoretical Sciences
for partial support.

\end{document}